\pdfoutput=1

\documentclass[11pt]{article}
\usepackage{hyperref} 
\usepackage[final]{acl}
\usepackage{float}
\usepackage{subcaption}
\usepackage{amsmath} 
\usepackage{times}
\usepackage{latexsym}

\usepackage[T1]{fontenc}

\usepackage[utf8]{inputenc}

\usepackage{microtype}

\usepackage{inconsolata}

\usepackage{graphicx}

\title{Unveiling Behavioral Differences in Bilingual Information Operations: A Network-Based Approach}

\author{Bowen Yi \\
  Department of Computer Science and Engineering,\\University of Michigan, Ann Arbor\\
  \texttt{bowenyi@umich.edu}}

\begin{document}
\maketitle

\begin{abstract}
Twitter has become a pivotal platform for conducting information operations (IOs), particularly during high-stakes political events. In this study, we analyze over a million tweets about the 2024 U.S. presidential election to explore an under-studied area: the behavioral differences of IO drivers from English- and Spanish-speaking communities. Using similarity graphs constructed from behavioral patterns, we identify IO drivers in both languages and evaluate the clustering quality of these graphs in an unsupervised setting. Our analysis demonstrates how different network dismantling strategies, such as node pruning and edge filtering, can impact clustering quality and the identification of coordinated IO drivers. We also reveal significant differences in the topics and political indicators between English and Spanish IO drivers. Additionally, we investigate bilingual users who post in both languages, systematically uncovering their distinct roles and behaviors compared to monolingual users. These findings underscore the importance of robust, culturally and linguistically adaptable IO detection methods to mitigate the risks of influence campaigns on social media. Our code and data are available on \href{https://github.com/bowenyi-pierre/humans-lab-hackathon-24}{GitHub}. 

\end{abstract}

\section{Introduction}
Social media has become a crucial platform for influence campaigns aimed at shaping public perception. With Twitter boasting over 335 million monthly active users\footnote{https://www.statista.com/statistics/303681/twitter-users-worldwide/}, these campaigns have the potential to significantly influence political discourse. Often backed by state actors, information operations (IOs) are particularly prevalent during critical geopolitical events such as the 2024 U.S. presidential election \citep{Minici2024UncoveringCC, Shah2024UnfilteredCA}. Ensuring fairness and transparency in political campaigns requires robust methods for detecting IO drivers on social media.

While research has highlighted the prevalence of IO drivers in lower-resourced languages like Spanish, their impact is no less severe than English-based campaigns \citep{Haider2020DetectingSM,Sharma2020IdentifyingCA}. This underscores the need to detect coordinated users in multilingual contexts. However, traditional machine learning methods often introduce biases against non-English communities, reducing detection accuracy \citep{gallegos2024biasfairnesslargelanguage, masud-etal-2024-hate}. Conversely, rule-based approaches struggle to scale given the immense volume of social media data.

Motivated by these challenges, we first analyze behavioral traces—such as shared domains and hashtags—that highlight divergences between English and Spanish users. Using these traces, we employ network-based methods to detect IO drivers. We introduce novel network dismantling techniques incorporating temporal and sentiment-based filters. Through a newly proposed evaluation framework for clustering quality, we demonstrate how these techniques affect English and Spanish data differently, emphasizing the need for language-specific IO detection methods. Applying the best-performing technique, we identify IO drivers and investigate behavioral differences between English and Spanish users during the 2024 U.S. presidential election. Understanding these differences provides valuable insights for future methods to detect coordinated malicious activities in multilingual settings. Finally, we explore the unique role of bilingual users who post in both languages, revealing their distinct behavioral strategies and their function as bridges between linguistic communities.

This study addresses the following research questions: \begin{itemize} \item \textbf{RQ1}: How do network-based methods for behavioral traces identify IO drivers in different languages? How do these methods, along with edge-weight filtering and node-pruning techniques, impact clustering quality in the absence of labeled data? \item \textbf{RQ2}: How do detected IO drivers differ in their behavior across languages, including their discussed topics, sentiment, political leanings, and media preferences? \item \textbf{RQ3}: How do bilingual posters differ in their behavior from monolingual posters, and what roles do they play in bridging linguistic communities? \end{itemize}

Finally, we acknowledge the limitations of our work, including constraints in computational resources and time, and suggest future research directions to build upon our findings. 

This work in progress is the result of a two-week research Hackathon hosted by HUMANS-Lab at the University of Southern California. Due to restrictions imposed by the event on permissible datasets, we were unable to use external data to validate the accuracy of IO driver detection in the current version. We plan to update the evaluation with additional datasets in future versions of this work.







\section{Related Work}

\subsection{IO Driver Detection} A growing body of research has focused on methods for detecting IO drivers. For instance, \citet{luceri2023unmaskingwebdeceituncovering} propose combining centrality-based node pruning with multiple similarity networks to create a fused network for accurate IO driver detection. This study builds on their work by evaluating how these techniques impact clustering quality, an aspect not addressed in the original paper. Additionally, the authors of \citep{luceri2023unmaskingwebdeceituncovering} leverage network embeddings to develop machine learning models for state-of-the-art IO driver detection. However, due to constraints in computational resources and the lack of labeled data, we did not explore this direction further. Furthermore, \citet{cinus2024exposingcrossplatformcoordinatedinauthentic} investigate cross-platform IOs spanning social media platforms such as Twitter and Telegram. 

It is notable that IO drivers could exist in multilingual settings, which increase the difficulty in detection. Most existing research on multilingual IO detection focuses on English, Russian, and Chinese. For example, \citet{8508646, yang2022botometer} document coordinated activities originating in Russia aimed at shaping public opinion during events like the 2016 U.S. election. \citet{Zannettou2018WhoLT} also found that Russian coordinated accounts have a high tendency to post in English.  Similarly, studies have uncovered IOs backed by the Chinese government \citep{10.1007/978-3-031-17114-7_9, jacobs2024whatisdemocracy}. In contrast, IO detection in Spanish—explored in this paper—has received relatively less attention, despite Spanish-language accounts backed by the Cuban and Venezuelan governments playing a significant role in spreading misinformation and shaping political opinions on social media \citep{wang2023evidence}. Furthermore, \citet{bogonez2023russian} demonstrated how Russian trolls strategically posted in Spanish to influence Spanish-speaking audiences during the Russia-Ukraine war.

\citet{Hui2022CoordinatedTA, 10.1145/3366424.3385775, Pacheco2020UncoveringCN} systematically analyze how different behavioral traces—such as performing synchronized activities and disseminating identical images—can be incorporated into unsupervised, network-based methodologies to uncover coordinated account groups. These studies inspired our edge-filtering strategies, which incorporate short time windows (as described in Section \ref{sec: network fusion}), as well as constructing Co-Domain, Co-Hashtag, and Text Similarity networks.

\subsection{Bilingual Users on Social Media} Previous research has demonstrated that bilingual users on social media exhibit distinct behavioral patterns compared to monolingual users. For example, \citet{Gavilanes2015LanguageTA} observed that the number of languages one speaks affects the interaction dynamics on Twitter: multi-lingual users are more likely to interact with others than monolingual users. Additionally, \citet{Mendelsohn2023BridgingNQ} examine the role of bilingual users in bridging online social networks using causal inference techniques, a promising avenue for future research.   

\section{Data}
This paper is based on the Twitter dataset on the 2024 U.S. Presidential Election introduced by \citet{balasubramanian2024publicdatasettrackingsocial}, which contains 22 million tweets. However, the dataset is predominantly English, with Spanish being the second most represented language. To avoid confounders caused by the differing data sizes when comparing the behaviors of English and Spanish IOs, we downsampled the English dataset to match the size of the Spanish data. Specifically, we randomly sampled an equal number of English tweets corresponding to the timeframe of the Spanish tweets. This process resulted in a balanced dataset comprising 588,839 tweets each for English and Spanish, produced by 227,414 and 125,909 unique users, respectively. In Table~\ref{tab:popular-tags}, we analyzed and compared the top 10 most popular URL domains and hashtags in English and Spanish tweets. To facilitate a direct comparison, we controlled for potential confounders such as dataset size. However, completely eliminating confounding factors remains challenging.

We identified 113,855 hashtags in English tweets and 168,281 in Spanish tweets, with 22,319 and 21,329 unique hashtags, respectively. To extract web domains, we utilized the tldextract library\footnote{https://pypi.org/project/tldextract/}, converting each URL into its base domain. In total, English tweets contained 60,534 shared web domains (5,847 unique), while Spanish tweets had 128,981 shared web domains (6,543 unique). Inspired by \citet{cinus2024exposingcrossplatformcoordinatedinauthentic}, we not only identified the five most popular domains but also analyzed their factuality and political leaning using \href{https://mediabiasfactcheck.com/}{Media Bias/Fact Check (MBFC)}.

Our analysis revealed that domains without associated leaning or factuality, such as YouTube.com and X.com, were the most popular in both English and Spanish tweets. Therefore, we excluded these non-informative domains from Tables~\ref{tab:en-popular-domains} and~\ref{tab:es-popular-domains}. The original unfiltered data can be found in Tables~\ref{tab:en-popular-domains-unfiltered} and~\ref{tab:sp-popular-domains-unfiltered} in the Appendix. According to Table~\ref{tab:en-popular-domains}, media sources with right-leaning or lower factuality received more attention among English users. In contrast, Table~\ref{tab:es-popular-domains} shows that Spanish users primarily engaged with left-leaning and high-factuality media sources.

\begin{table}[]
\resizebox{\columnwidth}{!}{%
\begin{tabular}{|cc|cc|}
\hline
\multicolumn{2}{|c|}{English}                                    & \multicolumn{2}{c|}{Spanish}                                     \\ \hline
\multicolumn{1}{|c|}{Tag}                       & Frequency      & \multicolumn{1}{c|}{Tag}                       & Frequency       \\ \hline
\multicolumn{1}{|c|}{trump2024}                 & 10499          & \multicolumn{1}{c|}{biden}                     & 7163            \\ \hline
\multicolumn{1}{|c|}{maga}                      & 8441           & \multicolumn{1}{c|}{trump2024}                 & 6260            \\ \hline
\multicolumn{1}{|c|}{trump}                     & 4553           & \multicolumn{1}{c|}{trump}                     & 6194            \\ \hline
\multicolumn{1}{|c|}{biden}                     & 3382           & \multicolumn{1}{c|}{eeuu}                      & 4457            \\ \hline
\multicolumn{1}{|c|}{bidenharris2024}           & 1391           & \multicolumn{1}{c|}{maga}                      & 3211            \\ \hline
\multicolumn{1}{|c|}{donaldtrump}               & 1124           & \multicolumn{1}{c|}{mundo}                     & 2941            \\ \hline
\multicolumn{1}{|c|}{kamalaharris}              & 934            & \multicolumn{1}{c|}{donaldtrump}               & 2700            \\ \hline
\multicolumn{1}{|c|}{gop}                       & 863            & \multicolumn{1}{c|}{internacional}             & 2509            \\ \hline
\multicolumn{1}{|c|}{usa}                       & 734            & \multicolumn{1}{c|}{estadosunidos}             & 2252            \\ \hline
\multicolumn{1}{|c|}{joebiden}                  & 733            & \multicolumn{1}{c|}{usa}                       & 2159            \\ \hline 
\end{tabular}%
}
\caption{Top 10 most popular hashtags in English and Spanish tweets. All hashtags were lowercased to avoid duplicate counts. English tweets contained 113,855 hashtags (22,319 unique), while Spanish tweets had 168,281 hashtags (21,329 unique).}

\label{tab:popular-tags}
\end{table}

\begin{figure*}[ht]
    \centering
    \begin{subfigure}[t]{0.48\linewidth}
        \centering
        \resizebox{\linewidth}{!}{%
        \begin{tabular}{|c|c|c|c|}
        \hline
        Domain        & Frequency & Factuality & Leaning      \\ \hline
        foxnews.com   & 1887      & Mixed      & Right        \\ \hline
        breitbart.com & 1407      & Mixed      & Right        \\ \hline
        msn.com       & 1126      & High       & Left-Center  \\ \hline
        smartnews.com & 1118      & Mixed      & Least Biased \\ \hline
        newsbreak.com & 1118      & Mixed      & Left-Center  \\ \hline
        \end{tabular}
        }
        \caption{Top 5 most popular web domains in \textbf{English} tweets. There were 60,534 domains/URLs in English data.}
        \label{tab:en-popular-domains}
    \end{subfigure}
    \hfill
    \begin{subfigure}[t]{0.48\linewidth}
        \centering
        \resizebox{\linewidth}{!}{%
        \begin{tabular}{|c|c|c|c|}
        \hline
        Domain        & Frequency & Factuality & Leaning      \\ \hline
        elpais.com    & 2349      & High       & Left-Center  \\ \hline
        infobae.com   & 1220      & High       & Left-Center  \\ \hline
        rt.com        & 1088      & Very Low   & Right-Center \\ \hline
        clarin.com    & 1076      & High       & Right-Center \\ \hline
        univision.com & 1031      & High       & Left-Center  \\ \hline
        \end{tabular}
        }
        \caption{Top 5 most popular web domains in \textbf{Spanish} tweets. 128,981 domains/URLs were shared in Spanish data.}
        \label{tab:es-popular-domains}
    \end{subfigure}
    \caption{Comparison of the 5 most popular web domains in English and Spanish tweets. Factuality and political leaning were obtained from MBFC. Non-informative domains without MBFC records were excluded from analysis. English tweets contained more right-leaning and lower-factuality domains compared to Spanish tweets.}

    \label{fig:popular-domains}
\end{figure*}

\section{Experimentation}
\subsection{IO Driver Detection in English and Spanish Tweets}

This section focuses on addressing \textbf{RQ1}. It presents methodologies for detecting coordinated activities in English and Spanish tweets and analyzes the behavioral differences between their respective IO drivers. Coordinated accounts employ various strategies, such as sharing specific URLs, hashtags, or semantically similar tweets. Building on prior studies \citep{luceri2023unmaskingwebdeceituncovering, cinus2024exposingcrossplatformcoordinatedinauthentic, Pacheco2020UncoveringCN}, we identify relevant behavioral traces and construct similarity networks to capture these coordination patterns. In these networks, user similarities are represented as weighted edges, enabling the identification of coordinated groups.

Due to the lack of labeled data, we rely on unsupervised methods to detect IOs based on these similarity networks. Eventually, we construct a fused network that aggregates information from multiple behavioral traces. This fused network, enhanced by edge filtering and node pruning, has been shown to improve IO detection accuracy in previous studies \citep{luceri2023unmaskingwebdeceituncovering}. We use this fused network to identify the final set of IO drivers for subsequent analyses.

For most behavioral traces, we begin by forming a bipartite graph between users and entities, where entities represent the specific behavioral trace under consideration (e.g., web domains for the Co-Domain trace). Users are linked to entities based on their sharing activities, and edge weights are assigned using TF-IDF to reflect the importance of each entity. These bipartite graphs are then transformed into similarity networks, where connections between users are weighted by the cosine similarity of their TF-IDF vectors.

\begin{itemize}
\item \textbf{Co-Domain}: Captures user similarity based on shared web domains.
\item \textbf{Co-Hashtag}: Captures user similarity based on identical hashtag sequences.
\item \textbf{Text Similarity}: Captures user similarity based on semantically similar tweets.
\end{itemize}

Building these networks in later sections showed that English and Spanish data had different network structure. In addition to these behavioral traces, we explored Co-Mentioned Users (mentioning the same users in tweets) and Co-Retweet (resharing the same tweet). However, we excluded them from the analysis due to their limitations. Co-Mentioned Users can be noisy, as we notice that both organic users and IO drivers frequently mention the same influential accounts. Moreover, our dataset contains sparse retweet data, with less than 1\% of tweets being retweets, undermining its reliability for IO detection. 



\subsubsection{Co-Domain Network}
\label{sec:co-domain}

To address the sparse distribution of individual URLs, we converted each URL into its base domain. As observed in Tables~\ref{tab:en-popular-domains-unfiltered} and~\ref{tab:sp-popular-domains-unfiltered}, non-informative domains such as X.com frequently appeared. Consequently, we filtered out these domains during network construction. The full list of filtered domains can be found in Section~\ref{subsec:filtered-domains}.

Additionally, we filtered out users who had shared fewer than three unique web domains, resulting in 1,244 active users. After constructing a user-domain matrix, we applied a minimum document frequency of 3 to exclude rare domains. TF-IDF was then used to weigh domains, and we retained only edges with a cosine similarity above 0.6.

\begin{figure*}[ht!]
\centering
\begin{subfigure}[b]{0.48\linewidth}
\centering
\includegraphics[width=\linewidth]{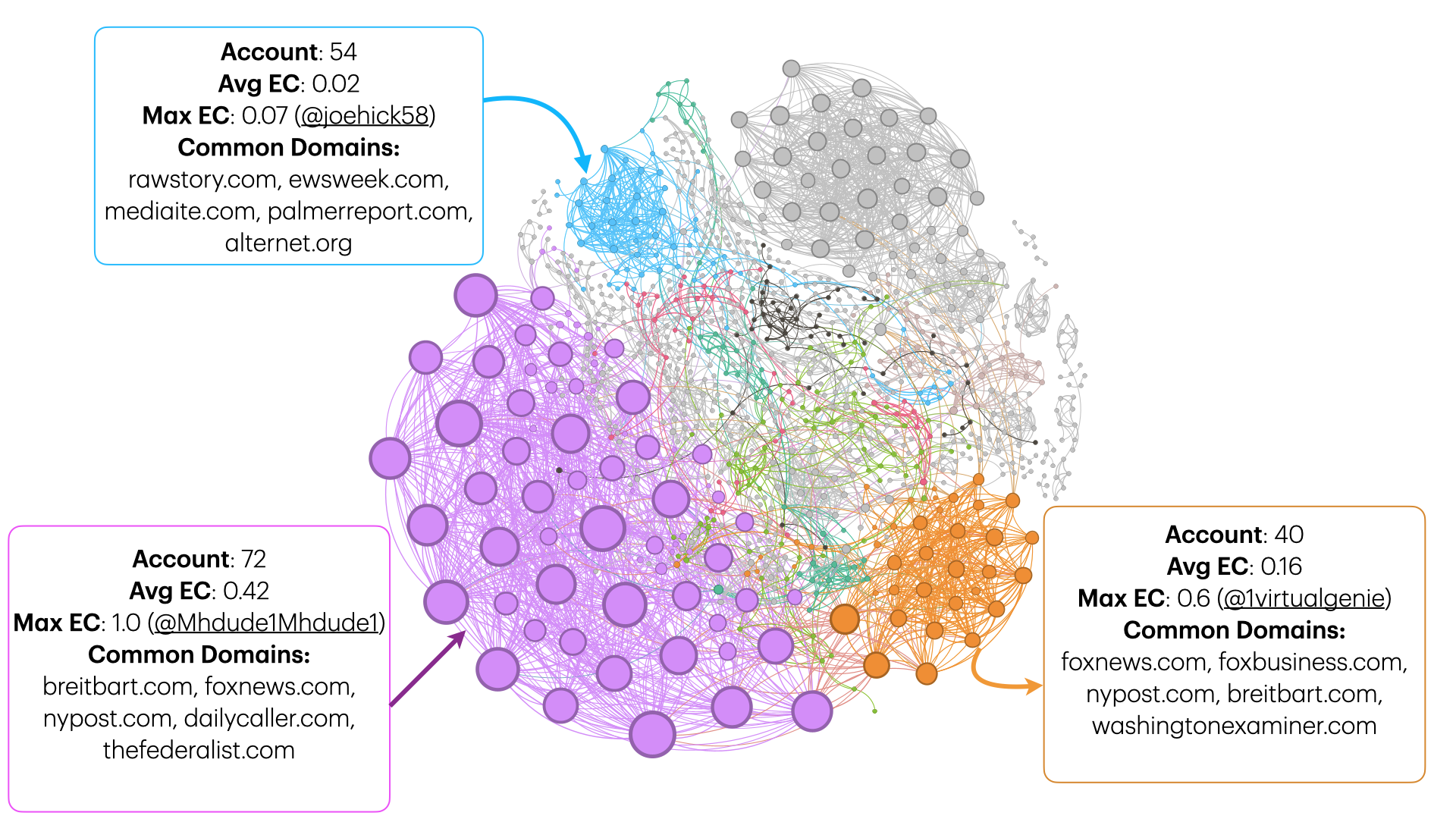}
\caption{English network: The cluster in purple has the highest average eigenvector centrality (Avg EC). Most large clusters feature a concentration of right-leaning domains.}
\label{fig:en-co-domain}
\end{subfigure}
\hfill
\begin{subfigure}[b]{0.48\linewidth}
\centering
\includegraphics[width=\linewidth]{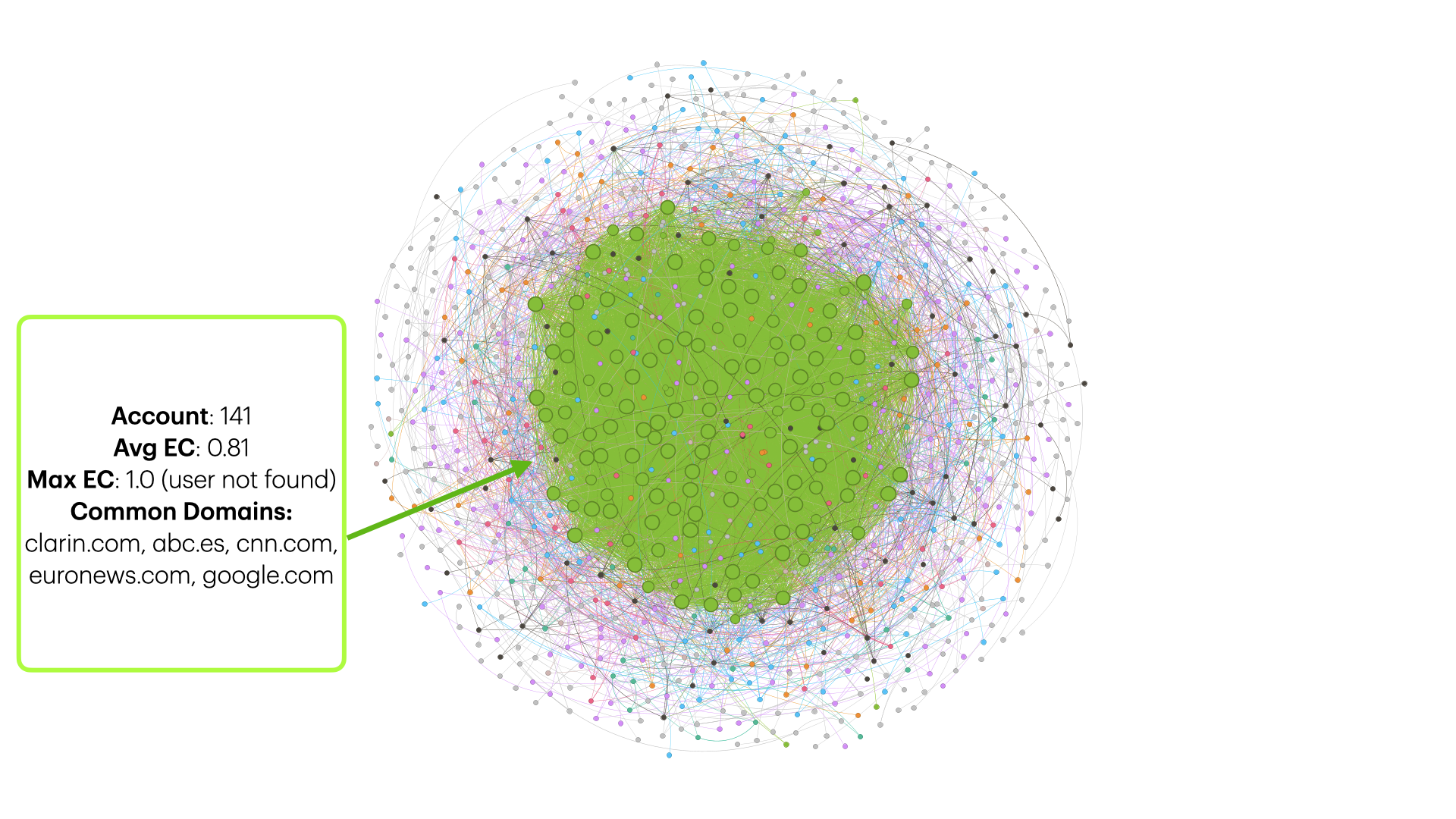}
\caption{Spanish network: The cluster in green has a significantly higher Avg EC than other clusters, featuring domains with more diverse political leanings.}
\label{fig:es-co-domain}
\end{subfigure}
\caption{Co-Domain networks for English (left) and Spanish (right) tweets. Each node represents a Twitter account, with size proportional to the eigenvector centrality. Clusters are colored differently, and key clusters are annotated with their characteristics, including the most widely shared web domains.}
\label{fig:co-domain-networks}
\end{figure*}

The resulting English network comprised 922 nodes and 3,254 edges. Using the Louvain algorithm\footnote{Unless specified, we use this algorithm throughout the paper, with a resolution of 1.0.} \citep{blondel2008fast} in Gephi \citep{bastian2009gephi}, we identified 95 communities with a modularity of 0.827. The network visualization is shown in Figure~\ref{fig:en-co-domain}, where large clusters predominantly feature right-leaning domains.

In contrast, the Spanish Co-Domain network exhibited denser connectivity, comprising 944 nodes and 9,625 edges. This indicates a greater overlap in shared domains among Spanish users compared to English users. We identified 79 communities with a modularity of 0.375. Notably, the second-largest cluster, consisting of 141 users, had a significantly higher average eigenvector centrality (EC) than any other cluster. Figure~\ref{fig:es-co-domain} illustrates the characteristics of this cluster, highlighting the diversity in political leaning among the shared domains.

\subsubsection{Co-Hashtag Network}
To preprocess the data, we filtered out users who shared fewer than six unique hashtags. We then applied a minimum document frequency (\texttt{min\_df} = 5) for TF-IDF weighting and retained edges with a cosine similarity threshold of 0.7.

\begin{figure*}[] \centering \begin{subfigure}[b]{0.48\linewidth} \centering \includegraphics[width=\linewidth]{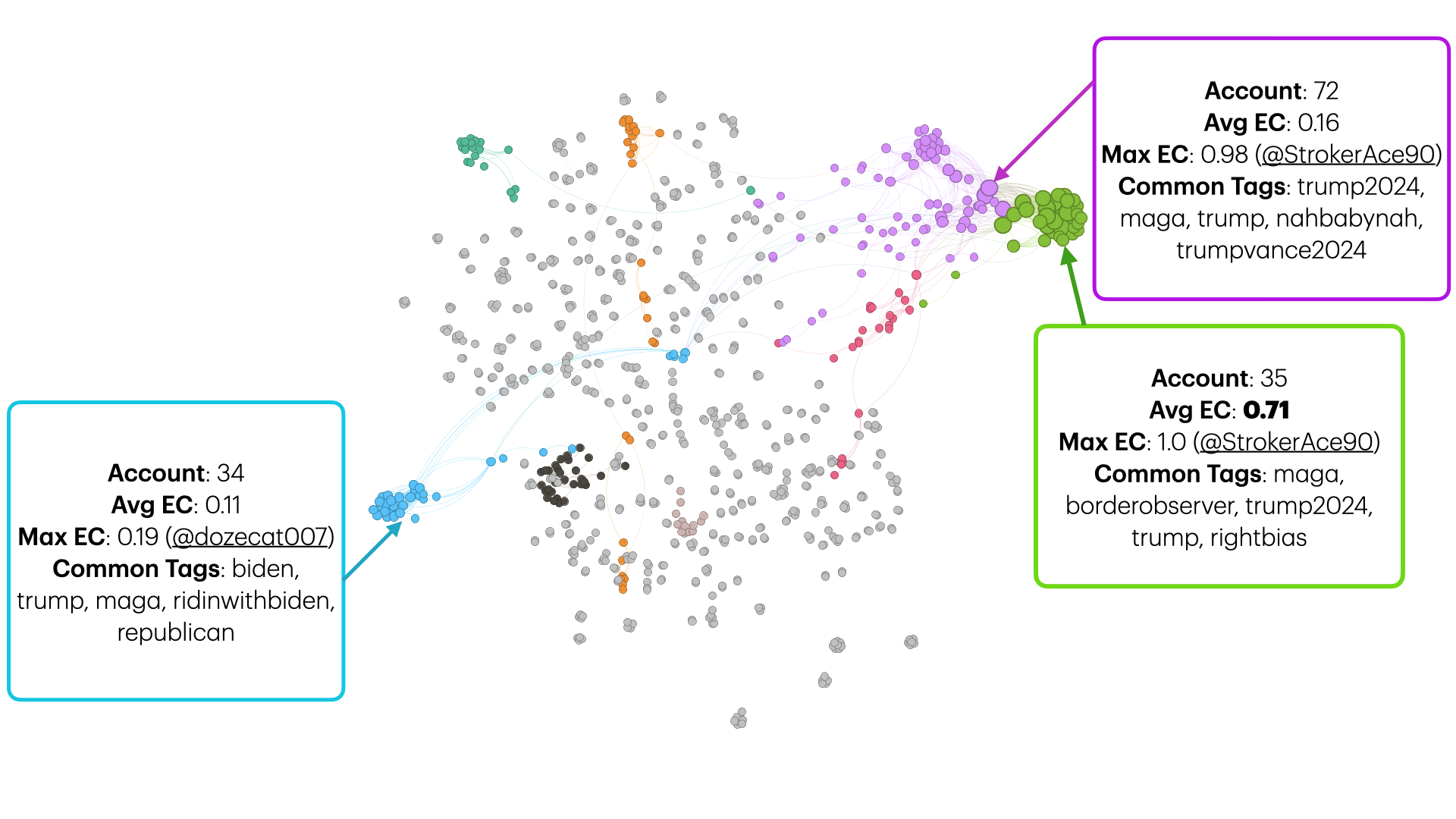} \caption{English network: The third-largest cluster (blue) shows a higher frequency of Biden-related tags, suggesting left-leaning tendencies. Other clusters frequently feature Trump-related tags.} \label{fig:en-co-tag} \end{subfigure} \hfill \begin{subfigure}[b]{0.48\linewidth} \centering \includegraphics[width=\linewidth]{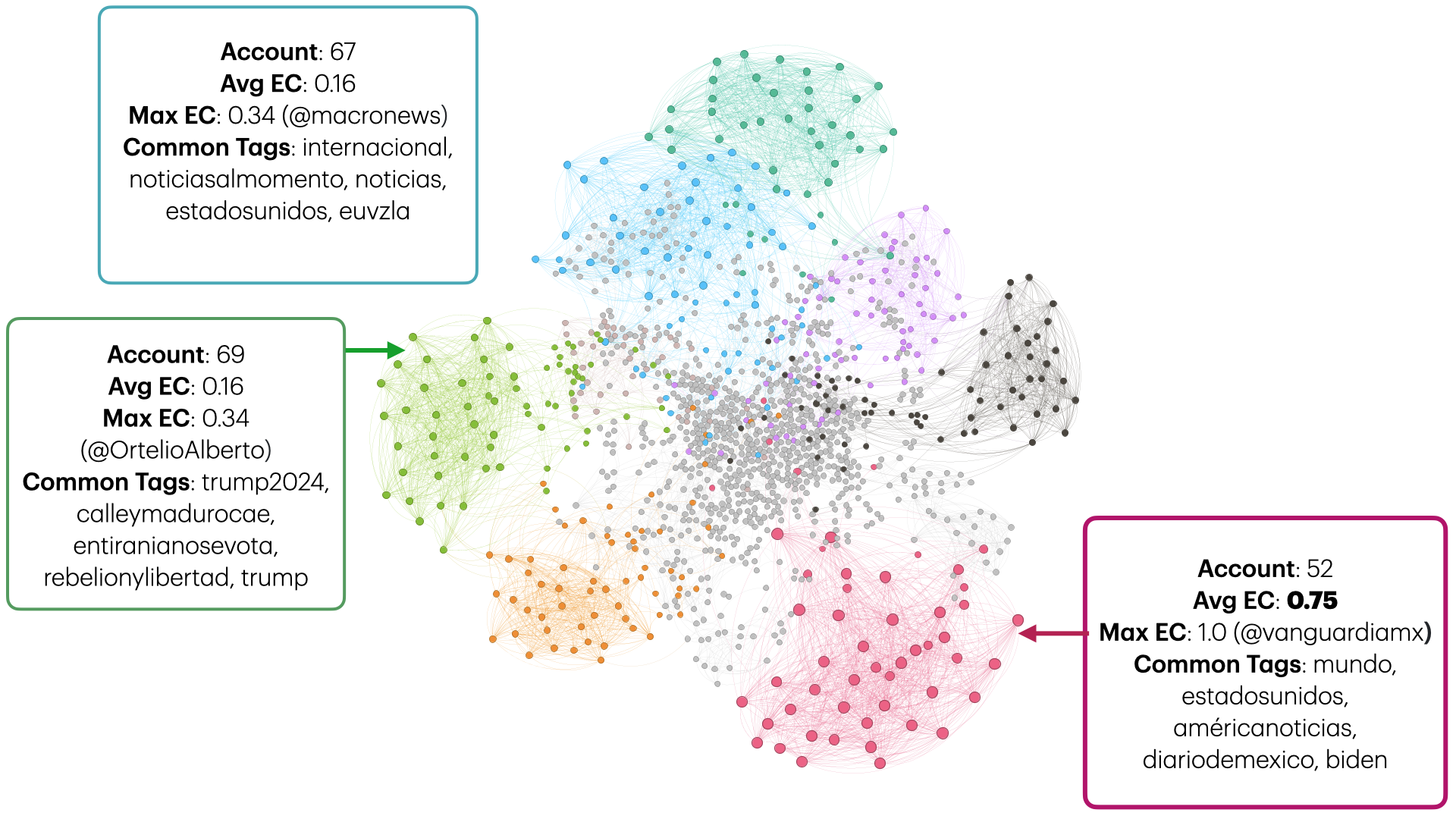} \caption{Spanish network: The largest cluster (purple) has low average eigenvector centrality (Avg EC), while the green cluster is potentially right-leaning due to frequent Trump-related tags.} \label{fig:es-co-tag} \end{subfigure} \caption{Co-Hashtag networks for English (left) and Spanish (right) tweets. Node sizes are proportional to eigenvector centrality. Key clusters are analyzed, including their five most common tags.} \label{fig:co-hashtag-networks} \end{figure*}

The resulting English network (Figure~\ref{fig:en-co-tag}) comprises 906 nodes and 1,865 edges. We detected 230 clusters with a modularity score of 0.833. Interestingly, user @StrokerAce90 had the maximum centrality across two of the largest clusters. Out of curiosity, we measured the botometer \cite{yang2022botometer} of this user, and found it had a low score (more details on this topic are included in Appendix \ref{subsec:high-centrality}). The Spanish network (Figure~\ref{fig:es-co-tag}) contains 1,262 nodes and 5,520 edges, with 205 clusters and a modularity score of 0.904. The high modularity in both networks suggests strong community structures, indicating minimal overlap in shared hashtags across user groups.

\subsubsection{Text Similarity Network}
We employed a similar preprocessing strategy to \cite{luceri2023unmaskingwebdeceituncovering} by removing punctuations, emojis, stopwords, and short tweets. Different from networks above, we build a direct similarity network instead of a bipartite graph. We converted each tweet into sentence embeddings using the sentence transformer distiluse-base-multilingual-cased-v1 from HuggingFace \citep{reimers-2019-sentence-bert}. Then, we calculated the cosine similarities of those embeddings, which serve as weights of the graph. Interestingly, we noticed that Spanish tweets had an overall higher text similarity, as setting the same similarity threshold results in a significantly bigger network for Spanish data. Therefore, we applied a higher similarity threshold = 0.95 for Spanish tweets, whereas 0.90 for English tweets. 

\begin{figure*}[]
    \centering
    \begin{subfigure}[b]{0.48\linewidth}
        \centering
        \includegraphics[width=\linewidth]{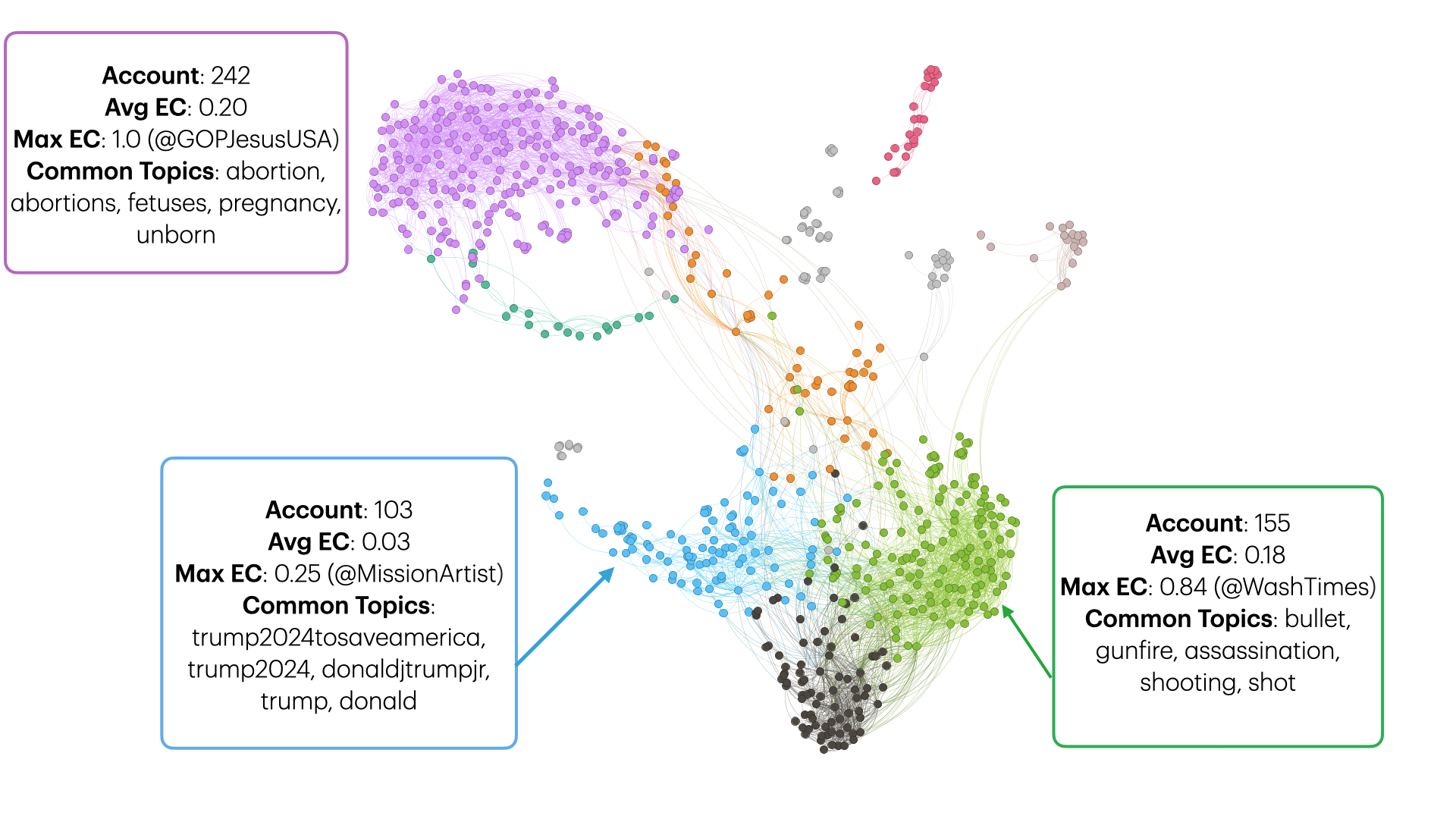}
        \caption{English network: The purple, green, and blue clusters consist of 32\%, 20\%, and 13\% of all nodes. The green cluster has the highest average EC. We witnessed a relatively high concentration of right-leaning topics in those clusters.}
        \label{fig:en-sim}
    \end{subfigure}
    \hfill
    \begin{subfigure}[b]{0.48\linewidth}
        \centering
        \includegraphics[width=\linewidth]{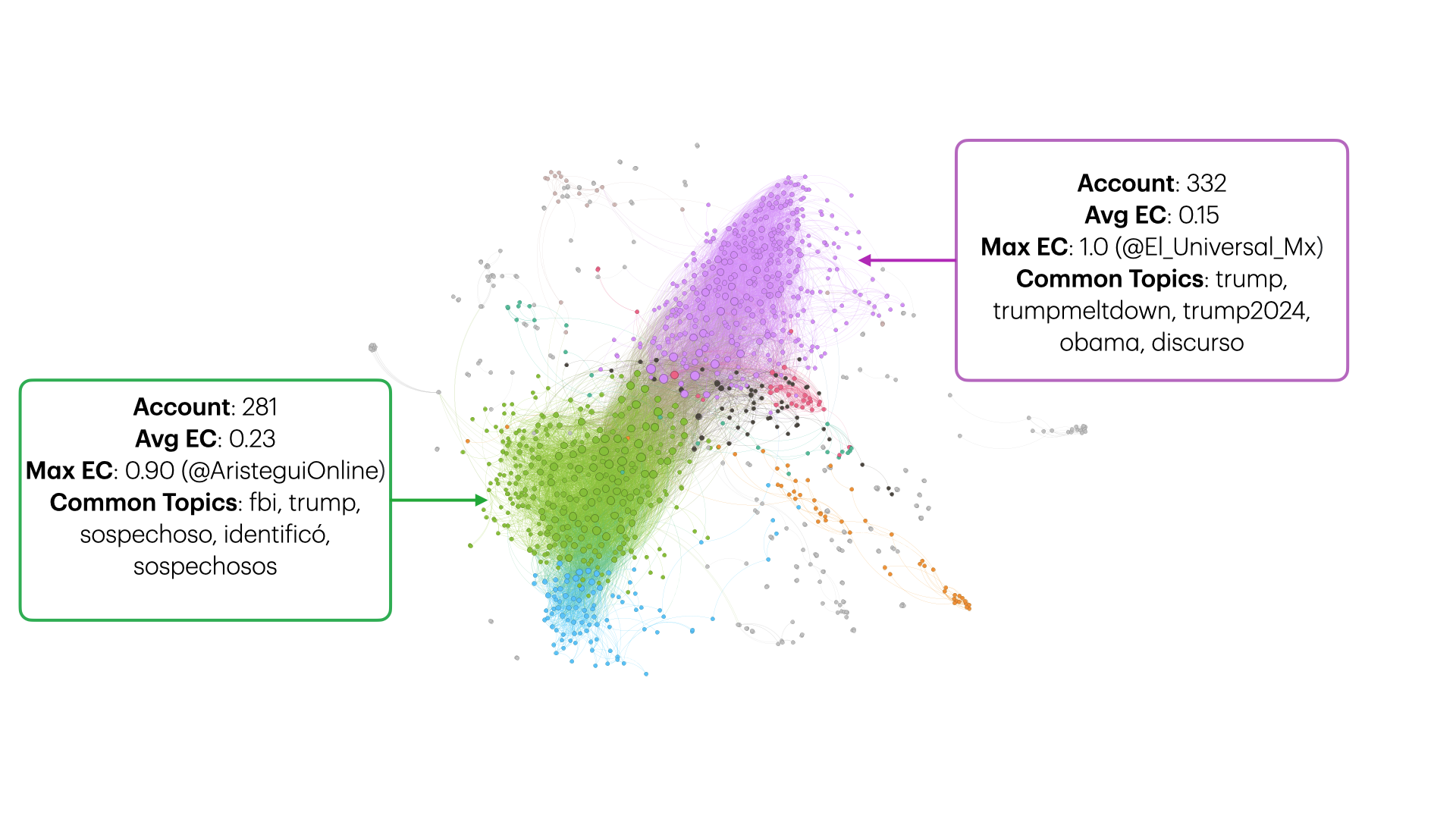}
        \caption{Spanish network: We illustrate the two biggest clusters identified, which cover 30\% and 25\% of all nodes respectively. The green cluster has the highest average EC. Similar to the English graph, we witnessed a high concentration of right-leaning topics in those clusters.}
        \label{fig:es-sim}
    \end{subfigure}
    \caption{Text similarity networks for English (left) and Spanish (right) tweets. Except for common cluster characteristics, we illustrate the 5 most common topics for representative clusters}
    \label{fig:text-sim-networks}
\end{figure*}

The resulting English network has 754 nodes and 4,058 edges. We detected 25 clusters with a modularity = 0.635. We visualize the network in Fig \ref{fig:en-sim} and analyzed the prevalent topics in selected clusters using BERTopic\footnote{Unless specified, we used this tool to obtain top topics.} \citep{grootendorst2022bertopic} with KeyBERTInspired as the representation model.  The Spanish network has 1,100 nodes with 14,280 edges. We identified 80 clusters with 0.442 modularity. We visualize the network in Fig \ref{fig:es-sim}. By comparing common topics in several bigger clusters from English and Spanish networks, we observed a high concentration of right-leaning topics.  

\subsubsection{Network Fusion}
\label{sec: network fusion}
Relying on a single behavioral trace often fails to capture the full range of coordinated activities, as noted by \citet{luceri2023unmaskingwebdeceituncovering}. To address this, we constructed a fused network combining the Co-Domain, Co-Hashtag, and Text Similarity networks. Fused networks have consistently demonstrated superior performance in identifying IO drivers compared to individual behavioral trace networks. Thus, the IO drivers identified in the fused network form the basis for our subsequent research questions. To enhance the fused network, we explored two complementary strategies: edge filtering and node pruning.

\paragraph{Edge Filtering} Edge filtering based on low weights is commonly used to uncover coordinated activities \citep{burghardt2023sociolinguisticcharacteristicscoordinatedinauthentic, suresh2023trackingfringecoordinatedactivity}. Additionally, synchronous activities within a time window \citep{Pacheco2020UncoveringCN, keller2020political} and shared sentiment \citep{alvarez2015sentiment, Xia2020SpreadOT} are distinguishing features of IO drivers. Inspired by these findings, we incorporated weight, time, and sentiment-based edge filtering to refine the network.

\paragraph{Node Pruning} In parallel, we applied node pruning based on eigenvector centrality, setting a threshold of $10^{-2}$. Node pruning has been shown to outperform edge filtering in \citep{luceri2023unmaskingwebdeceituncovering}, and we adopted this approach to ensure the retention of key coordinated accounts. Although our experiments showed that temporal and sentiment-based edge filtering produced dense networks with low modularity, the potential of these methods warrants further exploration (see Appendix~\ref{sec: edge filtering}).

Despite the lack of labeled data to directly evaluate IO detection accuracy, we conducted an extensive analysis of how different edge-filtering and node-pruning strategies affect clustering quality, as detailed in Section~\ref{subsec:evaluate}. Ultimately, we used high-centrality nodes from the fused network as the final set of IO drivers.  


\subsubsection{Evaluating Clustering Quality}
\label{subsec:evaluate}

This section introduces a novel method to evaluate the clustering quality of a network, focusing on the homogeneity of topics within clusters and the heterogeneity of topics across clusters. Unlike standard accuracy evaluations, our approach is designed for unsupervised settings without ground truth labels.  

To begin, we identified common topics in each cluster using BERTopic. Cluster homogeneity was measured using normalized entropy, inspired by \citet{rosenberg2007v}, where lower entropy indicates greater homogeneity. We then computed the Jensen-Shannon Divergence (JSD) \citep{fuglede2004jensen} between every pair of clusters to assess inter-cluster separation, with higher average JSD reflecting better separation.  

For each cluster \(k\), the entropy \(H_k\) was calculated as:  
\[
H_k = -\sum_{t \in T_k} p_{t,k} \frac{\log(p_{t,k})}{\log(|T_k|)}
\]
where \(T_k\) represents the set of topics in cluster \(k\), and \(p_{t,k}\) denotes the proportion of users in cluster \(k\) associated with topic \(t\). Given that cluster sizes vary, we computed a weighted average entropy:  
\[
H_{\text{weighted}} = -\frac{1}{N} \sum_{k=1}^K |C_k| \cdot H_k
\]
where \(N = \sum_{k=1}^K |C_k|\) is the total number of nodes, and \(|C_k|\) is the size of cluster \(k\).  

\begin{figure*}[]
    \centering
    \begin{subfigure}[b]{0.48\linewidth}
        \centering
        \includegraphics[width=\linewidth]{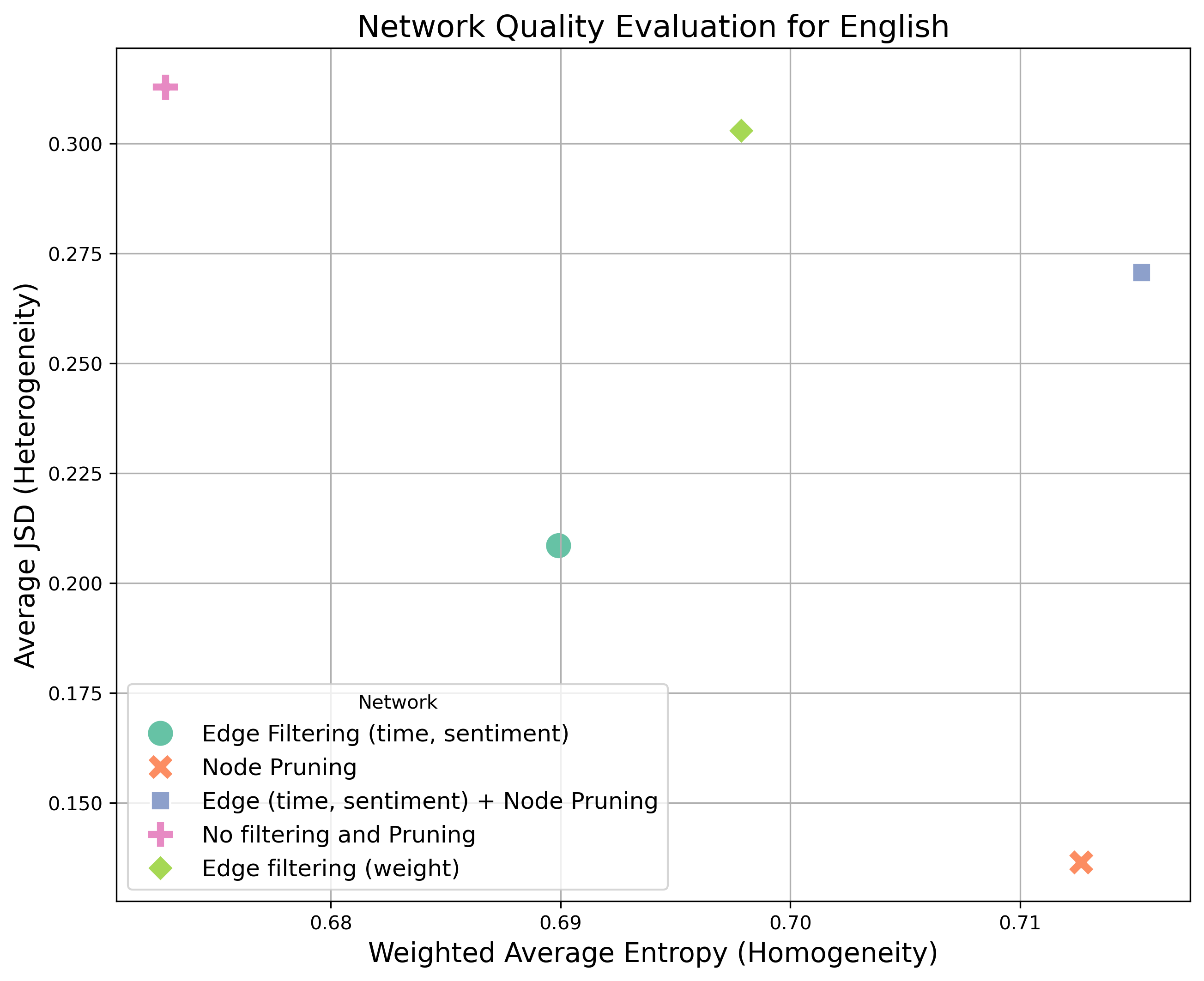}
        \caption{English network: No filtering or pruning yielded the highest clustering quality. Overall, none of the methods produced high-quality clusters.}
        \label{fig:en-eval}
    \end{subfigure}
    \hfill
    \begin{subfigure}[b]{0.48\linewidth}
        \centering
        \includegraphics[width=\linewidth]{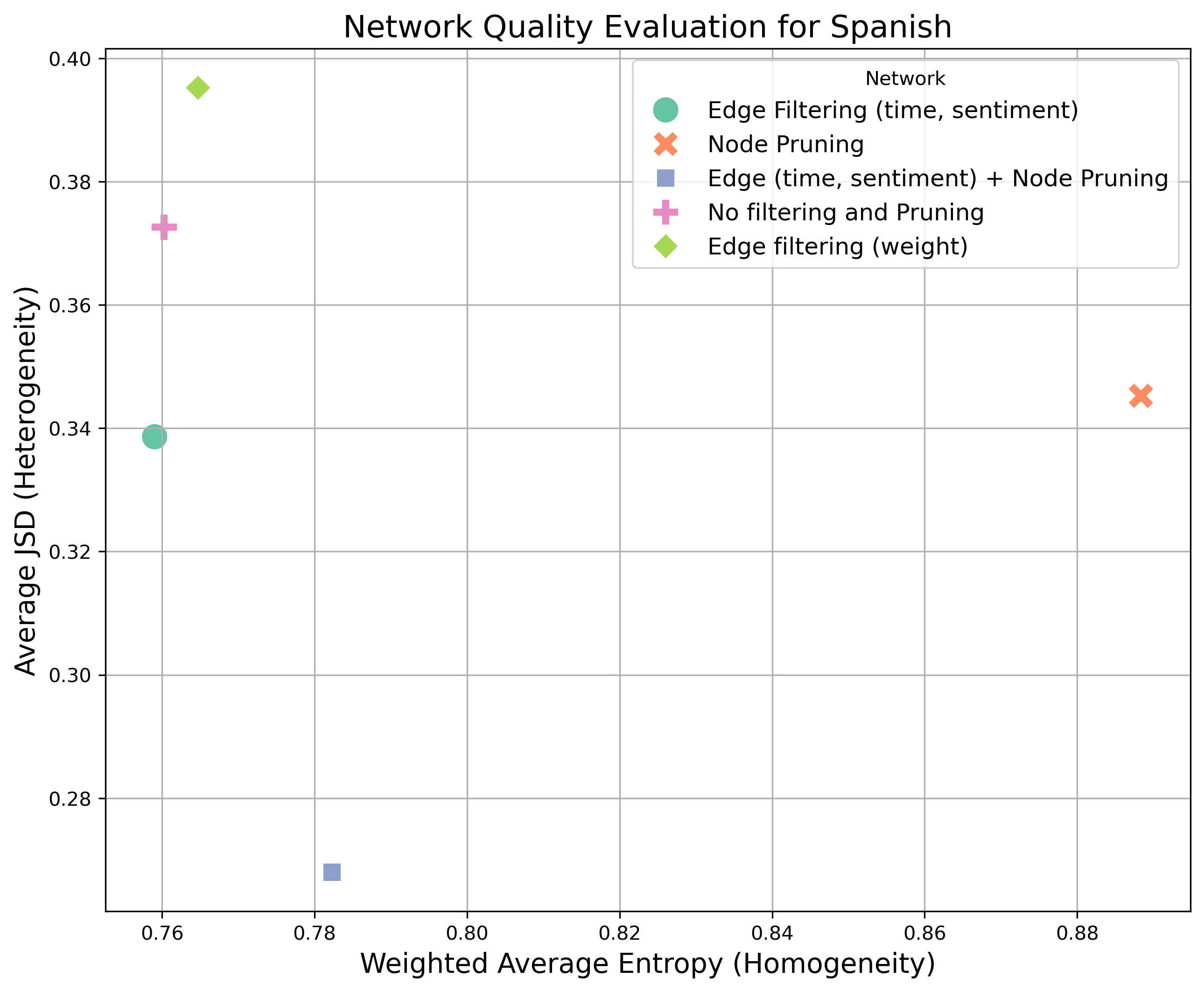}
        \caption{Spanish network: Weight-based edge filtering produced the best clustering quality, though overall results were suboptimal.}
        \label{fig:es-eval}
    \end{subfigure}
    \caption{Clustering quality evaluation for English (left) and Spanish (right) fused networks after five different operations. Small clusters without topics were excluded from the analysis. For weight-based edge filtering, only the top 30\% of edges were retained.}
    \label{fig:en-sp-eval}
\end{figure*}

We plotted \(H_{\text{weighted}}\) on the x-axis and average JSD on the y-axis (Figure~\ref{fig:en-sp-eval}). Ideally, a high-quality network would appear in the top-left quadrant, indicating low intra-cluster entropy and high inter-cluster separation. However, none of the fused networks achieved satisfactory clustering quality, as evidenced by high \(H_{\text{weighted}}\) and low JSD values. Notably, edge filtering and node pruning strategies appeared to degrade clustering quality, particularly for English data.  

We attribute these results to two factors: (1) edge filtering and node pruning may have excessively reduced the number of (weak but meaningful) edges, and (2) clustering algorithms and parameters likely require further tuning. While clustering quality (e.g., modularity) does not directly correlate with IO detection performance, these findings highlight the need for language-specific optimization of edge filtering and node pruning parameters. Figures~\ref{fig:en-eval} and~\ref{fig:es-eval} illustrate the stark differences in clustering quality between English and Spanish networks despite using identical parameters. Future research should refine these methods to improve clustering quality while maintaining IO detection accuracy.


\subsection{Differences between English and Spanish IO Drivers}
\label{subsec:en-sp-driver-diff}
This section addresses \textbf{RQ2}. It highlights the behavioral differences between English and Spanish IO drivers in terms of temporal topic shifts, sentiment, and engagement patterns (likes, replies, quotes, and retweets). The IO drivers are identified based on the findings in Section~\ref{subsec:evaluate}. For English data, IO drivers are defined as the top 3\% of nodes by eigenvector centrality in the fused network without any filtering or pruning. For Spanish data, IO drivers are selected as the top 5\% of nodes by centrality from the fused network after filtering out the bottom 70\% of low-weighted edges. The differing percentile criteria account for the disparity in graph sizes between the two languages. This process resulted in 72 English IO drivers and 61 Spanish IO drivers.

Tables~\ref{tab:english_tags_domains} and~\ref{tab:spanish_tags_domains} summarize the top tags and domains used by English and Spanish IO drivers, respectively. Engagement statistics, including average likes, replies, quotes, and retweets, are presented in Table~\ref{tab:engagement}. Additionally, although not shown in this paper, we analyzed the monthly topical trends of IO drivers. While U.S. political figures dominate the discussions in both languages, Spanish IO drivers more frequently discuss foreign political figures and international affairs, such as Netanyahu and Putin. Topics like immigration and drug trafficking also receive greater attention among Spanish IO drivers. Further details on topical shifts can be found in the accompanying code repository.

Finally, we examined the sentiment shifts of IO drivers over time. As visualized in Figures~\ref{fig:en-sent-driver} and~\ref{fig:es-sent-driver} (Appendix~\ref{subsec:sentiment-io-drivers}), both English and Spanish IO drivers exhibit a similar amount of positive and negative sentiment, except during July and August 2024 when positive sentiment clearly outweigh negative sentiment. Future research could investigate the relationship between temporal sentiment shifts and political events.
 
\begin{table}[]
\centering
\resizebox{\columnwidth}{!}{%
\begin{tabular}{|c|c|}
\hline
\textbf{Top Tags (Frequency)} & \textbf{Top Domains (Frequency)} \\ \hline
trump2024 (317) & breitbart.com (82) \\ \hline
biden (61) & rawstory.com (63) \\ \hline
trump (49) & newsweek.com (46) \\ \hline
breakingnews (45) & washingtontimes.com (38) \\ \hline
nahbabynah (29) & foxnews.com (34) \\ \hline
\end{tabular}%
}
\caption{Top tags and domains for English IO drivers. This table reflects a strong focus on U.S. politics, with hashtags such as \textit{trump2024} and \textit{biden}, and domains like \textit{breitbart.com} and \textit{foxnews.com} that are often associated with conservative or right-leaning political perspectives. The presence of more politically charged content suggests a polarized political discourse.}
\label{tab:english_tags_domains}
\end{table}

\begin{table}[h!]
\centering
\resizebox{\columnwidth}{!}{%
\begin{tabular}{|c|c|}
\hline
\textbf{Top Tags (Frequency)} & \textbf{Top Domains (Frequency)} \\ \hline
últimahora (213) & infobae.com (510) \\ \hline
eu (159) & cnn.it (444) \\ \hline
esnoticia (71) & ntn24.com (313) \\ \hline
trump (68) & efe.com (293) \\ \hline
debate2024 (62) & lopezdoriga.com (271) \\ \hline
\end{tabular}%
}
\caption{Top tags and domains for Spanish IO drivers. Unlike the English table, this table focuses more on general news and current events, as reflected in hashtags such as \textit{últimahora} (breaking news) and \textit{esnoticia} (it’s news). Domains like \textit{infobae.com} and \textit{cnn.it} lean towards neutral or slightly left-leaning sources, indicating a different tone and focus compared to English tweets, with less emphasis on highly polarized political narratives.}
\label{tab:spanish_tags_domains}
\end{table}

\begin{table}[h!]
\centering
\resizebox{\columnwidth}{!}{%
\begin{tabular}{|c|c|c|c|c|}
\hline
        & Likes & Retweets & Quotes & Replies \\ \hline
English & 28.77 & 7.28     & 0.67   & 6.32    \\ \hline
Spanish & 28.37 & 8.56     & 1.05   & 5.33    \\ \hline
\end{tabular}
}
\caption{Engagement stats between English and Spanish IO drivers. We average the amount of all engagement metrics like likes and retweets. We see a similar pattern of engagement. }
\label{tab:engagement}
\end{table}

\subsection{Bilingual vs Monolingual Posters}

This section focuses on \textbf{RQ3}. It examines how bilingual posters—users who post in both English and Spanish—differ from monolingual posters in their behavior. We also investigate how these bilingual users behave differently when posting in one language versus the other. Additionally, we took a step into exploring the role of bilingual users in bridging English- and Spanish-speaking communities and their potential impact on information coordination across linguistic divides.


We identified 13,864 bilingual users who posted a total of 85,069 English tweets and 63,006 Spanish tweets. Table~\ref{tab:top_topics} presents the popular topics in English and Spanish tweets by bilingual posters. Notably, these users posted more content related to the Democratic Party in their English tweets compared to their Spanish tweets, which contrasts with the overall trend of right-leaning topics dominating English tweets. However, international events, such as the Russia-Ukraine war and the Israel-Palestine conflict, were prominent in both languages. Spanish tweets also displayed more polarized topics, often favoring a single political party.

In terms of engagement, bilingual users received higher average likes and retweets for their English tweets compared to their Spanish tweets. Tables~\ref{tab:top_domains_bilingual} and~\ref{tab:top_tags_bilingual} summarize the most popular web domains and hashtags shared by bilingual posters.

\begin{figure}
    \centering
    \includegraphics[width=\linewidth]{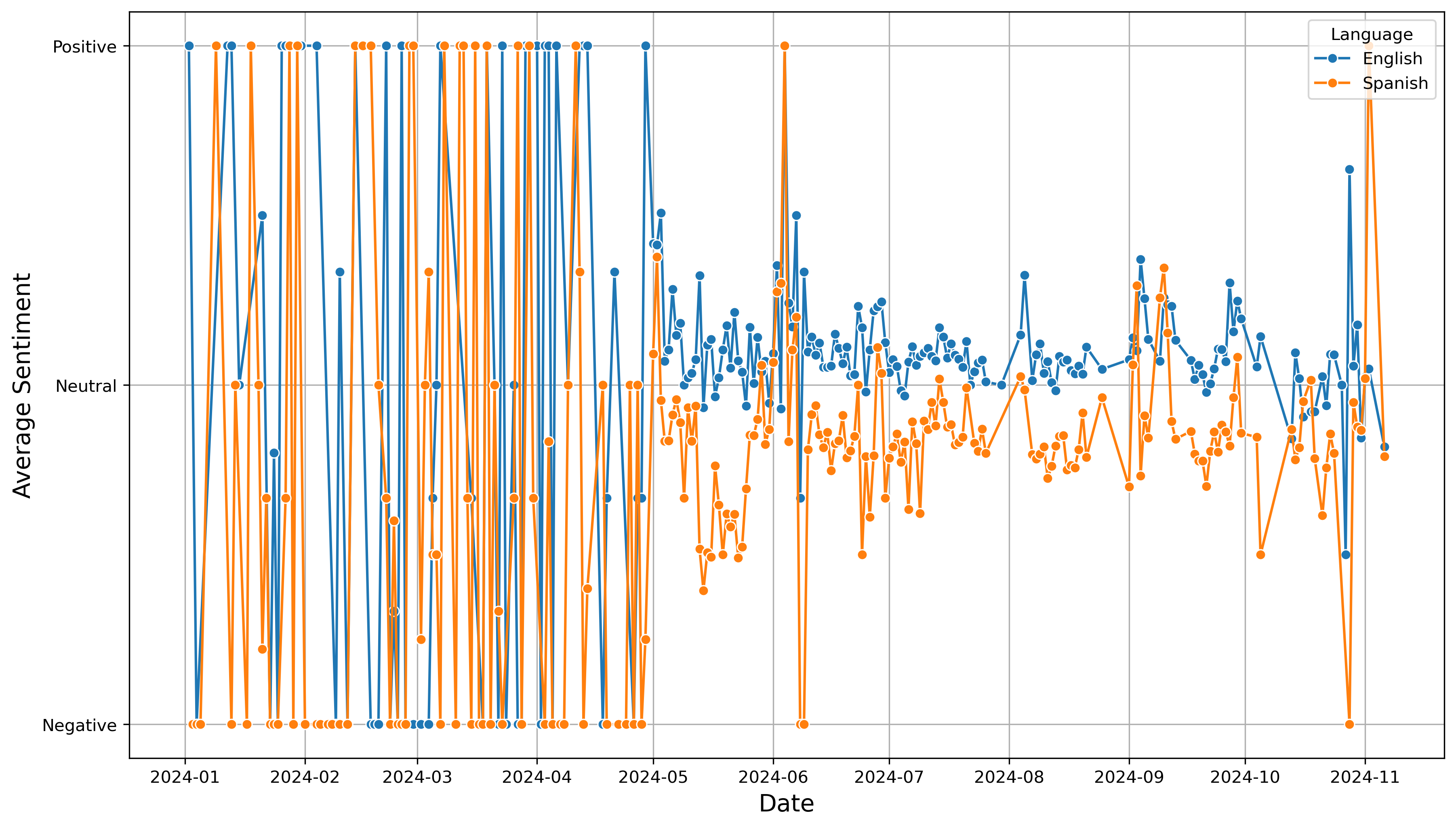}
    \caption{Sentiment shift of English and Spanish tweets posted by bilingual posters. Tweets posted outside year 2024 are filtered out.}
    \label{fig:sentiment}
\end{figure}

We also analyzed sentiment shifts over time, as illustrated in Figure~\ref{fig:sentiment}. While the overall sentiment trends were similar, English tweets displayed more positive sentiment, particularly in mid-to-late 2024. On average, English tweets had slightly higher sentiment scores than Spanish tweets, irrespective of time.

Finally, we examined the overlap between bilingual posters and the IO drivers identified in Section~\ref{sec: network fusion}. Interestingly, 53 English IO drivers were bilingual, whereas none of the Spanish IO drivers were. Assuming the IO drivers were accurately identified, this suggests that bilingual posting is not inherently indicative of coordinated behavior. Using bilingual status as a feature for detecting coordinated activities could lead to inaccurate or harmful results.

In conclusion, bilingual posters exhibit distinct behaviors when engaging with English and Spanish communities. Their topics of discussion and political leanings often differ from those of monolingual users. For instance, Republican-related content attracts more attention in their Spanish tweets, which contrasts with the findings in Table~\ref{tab:popular-tags}. Future research could explore the events that trigger code-switching on an individual level, the narratives of bilingual posters in other languages, and their role in bridging linguistic communities through a network science perspective.    

\begin{table}[h!]
\centering
\resizebox{\columnwidth}{!}{%
\begin{tabular}{|c|c|c|c|}
\hline
\textbf{Topic} & \textbf{Frequency} & \textbf{Likes} & \textbf{Retweets}\\ \hline
\multicolumn{4}{|c|}{\textbf{English Tweets}} \\ \hline
 U.S. Democratic Party & 3566 & 526.37 & 101.03 \\ \hline
 Democratic Female Politicians & 2857 & 77.38 & 26.46 \\ \hline
 Biden & 1610 & 25.17 & 17.41 \\ \hline
 Racism & 1192 & 112.31 & 57.82 \\ \hline
 Russia-Ukraine Conflict & 952 & 10.79 & 10.16 \\ \hline
\multicolumn{4}{|c|}{\textbf{Spanish Tweets}} \\ \hline
Pro-Democrats & 788 & 6.32 & 1.74 \\ \hline
Pro-Israel & 779 & 8.81 & 2.53 \\ \hline
Trump and Biden & 584 & 3.03 & 8.21 \\ \hline
Republicans & 470 & 6.24 & 2.23 \\ \hline
Pro-Trump & 452 & 36.00 & 13.03 \\ \hline
\end{tabular}%
}
\caption{Top topics, frequencies, and engagement stats for English and Spanish Tweets by Bilingual Posters. Non-meaningful topics were excluded for clarity. Likes and retweets have been averaged. Average quotes and quotes can also be found in the code file. English tweets have more interpretable and higher-quality topics than Spanish tweets.}
\label{tab:top_topics}
\end{table}

\section{Conclusion}
This study analyzed over a million English and Spanish tweets to develop network-based models for bilingual IO driver detection, leveraging multiple behavioral traces indicative of coordinated activities. We introduced a novel framework for evaluating the clustering quality of networks without ground truth labels and demonstrated how network dismantling techniques, such as edge filtering and node pruning, can influence clustering quality. Based on these evaluations, we identified IO drivers in both English and Spanish datasets. Our analysis revealed significant behavioral differences between English and Spanish users and IO drivers, such as their topic preferences. Furthermore, we systematically examined the behaviors of bilingual users who posted in both languages, uncovering distinct strategies in their engagement with different linguistic communities. These findings underscore the need for robust IO detection methods that account for linguistic and cultural differences. Also, by showing limitations of common IO detection techniques, this study lays the groundwork for future research on human-centered multilingual IO detection systems.

\section*{Limitations and Future Work}
\paragraph{Dataset} The primary limitation of this study is the absence of a labeled dataset for IO drivers, which limited our ability to evaluate detection accuracy. Access to labeled data would allow for error analysis across languages and improve IO detection methods. While generative AI, such as GPT-4, could assist in annotating data, time and resource constraints prevented its application in this study. 

In addition, limited dataset size, particularly for Spanish tweets, restricts the generalizability of our findings. To maintain balance, we downsampled the English dataset to match the Spanish dataset size, but this approach may have excluded valuable insights. Additionally, the lack of data in languages like Russian and Chinese—commonly associated with IOs—prevented a broader exploration of cross-lingual IOs. Future work should incorporate more diverse datasets to capture a wider range of coordinated activities.

\paragraph{Network Construction and Evaluation} We relied on the Louvain algorithm with a default resolution of 1.0 for cluster detection. Although this facilitated direct comparison, it limited our exploration of other clustering algorithms and parameters. Future work should experiment with alternative clustering methods to better capture community structures. Additionally, the assumption that high-centrality nodes represent IO drivers may lead to inaccuracies in supervised settings. Validating this assumption with labeled data is crucial.

Our evaluation of clustering quality focused on intra-cluster homogeneity and inter-cluster separation. While this approach offers some insights, it is by no means comprehensive even for unsupervised settings. Future research should develop more robust evaluation metrics for network quality without ground truth labels. Future work could also validate and improve the the proposed evaluation method on more data. 

Last but not least, future work could explore the deployment of LLMs in detecting IO drivers. As seen in \citep{luceri2023unmaskingwebdeceituncovering}, it would be interesting to use network embeddings as input. However, either using ML or network-only methods, it would be important to increase the interpretability of those methods. For example, why a model clusters certain points into one cluster or mark them as IO drivers. Inspecting those models can decrease potential harm in downstream applications.

\paragraph{Edge Filtering and Node Pruning} Our exploration of edge filtering in the fused network, based on weight, time window, and shared sentiment, yielded unsatisfactory results, likely due to overly stringent parameters. Future studies should refine these parameters and compare the effectiveness of edge filtering and node pruning in IO detection. Moreover, determining similarity thresholds for edge filtering manually is challenging and could be improved through automated or adaptive methods.
\paragraph{Additional Features for IO Detection} Future work could explore incorporating linguistic and stylistic features, such as formality and toxicity, into network fusion to enhance detection accuracy. It would also be promising to further investigate the role of AI-generated content and identical image sharing in IO driver detection. Similarly, integrating retweet data, which was scarce in our dataset, could provide insights into how IO drivers propagate specific sentiments, topics, or misinformation.
\paragraph{Cross-Lingual and Cross-Cultural IO Detection} While this study focused on bilingual IOs as one form of cross-lingual coordination, future research should explore other forms of cross-lingual IOs, particularly involving languages beyond English and Spanish. Understanding these dynamics could offer a more comprehensive view of global IO strategies. In addition, we noted significant differences in English and Spanish user behaviors, regardless of them being IO drivers or not. However, due to numerous confounder variables, we were unable to validate if those differences originate from cultural differences. Future research inspecting this angle would be exciting.

\section*{Ethics Impact}
This study is dealing with social media data, which could be sensitive. As a result, we carefully use the public available twitter dataset and acknowledge the possibility of identifiable information of users and have paid extra care in handling that. We didn't release any twitter post or twitter post that reveal identifiable information in this paper. 

In addition, users should be aware of the toxic posts made by certain users, which might harm their mental health. This paper doesn't show examples of toxic tweets, but the authors encountered them while processing the data, so users should be cautious. 

Last but not least, since our models in detecting IO drivers are by not free of errors, social media platforms should be cautious in leveraging them on content moderation. There is chance that our model flag a group of organic users to involve in malicious online activities, which can cause troubles on these users. Outside imperfect accuracy, social media moderators should also be aware of the biases or harms from such models on certain user groups. For example, studies show that ML models tend to flag people who speak Ebonic as toxic posters \citep{sap-etal-2019-risk}.  

\section*{Acknowledgements} We express our sincere gratitude to Emilio Ferrara at the University of Southern California for hosting the event, which made this project possible. We also acknowledge the University of Michigan for providing essential computing resources. This research was conducted without external funding.

\bibliography{custom}

\appendix

\section*{Appendix}

\label{sec:appendix}

\subsection*{Popular web domains without filtering}
Below, we identified the 10 most popular web domains shared in English and Spanish tweets. Unlike Section \ref{sec:co-domain}, we did not filter out domains that do not have records on MBFC. We noted that most popular domains, especially in Spanish data, do not have records on MBFC. Among those domains that have records on MBFC, right-leaning media sources receive a higher attention among English users than Spanish users.  
 
\begin{table}[H]
\resizebox{\columnwidth}{!}{%
\begin{tabular}{|c|c|c|c|}
\hline
Domain        & Frequency & Factuality & Leaning      \\ \hline
youtu.be      & 5155      & NA         & NA           \\ \hline
x.com         & 3576      & NA         & NA           \\ \hline
youtube.com   & 3003      & NA         & NA           \\ \hline
dlvr.it       & 2258      & NA         & NA           \\ \hline
foxnews.com   & 1887      & Mixed      & Right        \\ \hline
breitbart.com & 1407      & Mixed      & Right        \\ \hline
trib.al       & 1341      & NA         & NA           \\ \hline
msn.com       & 1126      & High       & Left-Center  \\ \hline
smartnews.com & 1118      & Mixed      & Least Biased \\ \hline
newsbreak.com & 1118      & Mixed      & Left-Center  \\ \hline
\end{tabular}
}
\caption{Frequency, factuality, and leaning of top 10 most popular web domains in \textbf{English} tweets. Factuality and leaning are based on data from MBFC.}
\label{tab:en-popular-domains-unfiltered}
\end{table}

\begin{table}[H]
\resizebox{\columnwidth}{!}{%
\begin{tabular}{|c|c|c|c|}
\hline
Domain      & Frequency & Factuality & Leaning     \\ \hline
dlvr.it     & 7524      & NA         & NA          \\ \hline
youtu.be    & 6001      & NA         & NA          \\ \hline
youtube.com & 3882      & NA         & NA          \\ \hline
x.com       & 3355      & NA         & NA          \\ \hline
bit.ly      & 3128      & NA         & NA          \\ \hline
elpais.com  & 2349      & High       & Left-Center \\ \hline
buff.ly     & 2341      & NA         & NA          \\ \hline
ift.tt      & 2223      & NA         & NA          \\ \hline
ow.ly       & 1804      & NA         & NA          \\ \hline
tinyurl.com & 1755      & NA         & NA          \\ \hline
\end{tabular}
}
\caption{Frequency, factuality, and leaning of top 10 most popular web domains in \textbf{Spanish} tweets. Factuality and leaning are based on data from MBFC.}
\label{tab:sp-popular-domains-unfiltered}
\end{table}

\subsection*{Uninformative Web Domains Filtered}
\label{subsec:filtered-domains}
To increase the interpretability and informativeness of Co-domain network, we filtered out several non-informative URLs that are among the top 20 shared among English and Spanish tweets. 

For English data, we omitted youtu.be, x.com, youtube.com, dlvr.it, trib.al, ift.tt, tiktok.com, bit.ly, yahoo.com. 

For Spanish data, we ignored dlvr.it, youtu.be, youtube.com, x.com, bit.ly, buff.ly, ift.tt, ow.ly, tinyurl.com, short.gy, trib.al, acortar.link, uni.vi. 

\subsection*{Edge Filtering and Node Pruning in Fused Networks}
\label{sec: edge filtering}

Filtering edges with low weights is a widely used technique for detecting coordinated activities \citep{burghardt2023sociolinguisticcharacteristicscoordinatedinauthentic, suresh2023trackingfringecoordinatedactivity}. However, \citet{luceri2023unmaskingwebdeceituncovering} found that node pruning often performs better. To complement weight-based filtering, we explored two additional criteria: temporal and sentiment alignment.

Temporal alignment was based on the observation that IO drivers often engage in synchronous activities within short time windows \citep{Pacheco2020UncoveringCN, keller2017manipulate}. We set a conservative time window of one hour, ensuring that nodes were connected only if their associated tweets occurred within this timeframe. Sentiment alignment was motivated by prior studies showing that coordinated activities frequently exhibit shared sentiment \citep{alvarez2015sentiment, Jiang2023AnalyzingTD}. We filtered edges based on sentiment similarity using a multilingual distilbert-based model\footnote{https://huggingface.co/lxyuan/distilbert-base-multilingual-cased-sentiments-student/discussions/8} for zero-shot classification.

Despite these refinements, temporal and sentiment-based edge filtering produced dense networks with poor cluster structure. For example, a network with a one-hour time window and shared sentiment contained 160 nodes and 1,354 edges, yielding a low modularity score of 0.279 with four communities detected. Using only the time window resulted in an even lower modularity score of 0.262. We attribute these outcomes to stringent filtering parameters, which may require further tuning.

While our results highlight the challenges of edge filtering, future work could investigate its comparative performance against node pruning in IO detection. The interplay between temporal, sentiment-based, and weight-based filtering remains an open question, especially given the lack of ground truth data. 




\subsection*{Fused Network Properties}
As a continuation of Section \ref{sec: network fusion}, we listed the network properties for English and Spanish tweets in Table \ref{tab:fused-net-properties}. This shows that the same strategy may have different impact on network properties for English and Spanish data. The cause might be due to inherent structure difference in the social networks.  
\label{subsec:fused-network-properties}
\begin{table*}[]
\resizebox{\textwidth}{!}{%
\begin{tabular}{|c|cccc|cccc|}
\hline
                                                & \multicolumn{1}{l|}{Nodes} & \multicolumn{1}{l|}{Edges} & \multicolumn{1}{l|}{Clusters} & \multicolumn{1}{l|}{Modularity} & \multicolumn{1}{c|}{Nodes} & \multicolumn{1}{c|}{Edges} & \multicolumn{1}{c|}{Clusters} & Modularity \\ \hline
                                                & \multicolumn{4}{c|}{English}                                                                                              & \multicolumn{4}{c|}{Spanish}                                                                         \\ \hline
Edge filtering (time, sentiment) + node pruning & \multicolumn{1}{c|}{160}   & \multicolumn{1}{c|}{1354}  & \multicolumn{1}{c|}{4}        & 0.27                            & \multicolumn{1}{c|}{382}   & \multicolumn{1}{c|}{11398} & \multicolumn{1}{c|}{3}        & 0.37       \\ \hline
Node pruning only                               & \multicolumn{1}{c|}{216}   & \multicolumn{1}{c|}{2114}  & \multicolumn{1}{c|}{4}        & 0.48                            & \multicolumn{1}{c|}{145}   & \multicolumn{1}{c|}{7421}  & \multicolumn{1}{c|}{4}        & 0.08       \\ \hline
Edge filtering (time, sentiment) only           & \multicolumn{1}{c|}{978}   & \multicolumn{1}{c|}{3446}  & \multicolumn{1}{c|}{89}       & 0.62                            & \multicolumn{1}{c|}{2144}  & \multicolumn{1}{c|}{19264} & \multicolumn{1}{c|}{196}      & 0.61       \\ \hline
Edge filtering on low weight only               & \multicolumn{1}{c|}{878}   & \multicolumn{1}{c|}{2745}  & \multicolumn{1}{c|}{116}      & 0.76                            & \multicolumn{1}{c|}{1223}  & \multicolumn{1}{c|}{8717}  & \multicolumn{1}{c|}{177}      & 0.57       \\ \hline
No filtering or pruning                         & \multicolumn{1}{c|}{2423}  & \multicolumn{1}{c|}{9150}  & \multicolumn{1}{c|}{291}      & 0.79                            & \multicolumn{1}{c|}{2867}  & \multicolumn{1}{c|}{29057} & \multicolumn{1}{c|}{255}      & 0.712      \\ \hline
\end{tabular}
}
\caption{Properties for English and Spanish fused networks after different node pruning and edge filtering techniques. Properties include number of nodes, edges, identified clusters, and modularity.}
\label{tab:fused-net-properties}
\end{table*}

\subsection*{Sentiment Shift for IO Drivers}
Below, we visualized the sentiment shift over time for 72 English IO drivers and 61 Spanish IO drivers. IO drivers from both languages exhibit a similar sentimental trend. Neutral sentiment is always the lowest type of sentiment.
\label{subsec:sentiment-io-drivers}
\begin{figure}[H]
    \centering
    \includegraphics[width=\linewidth]{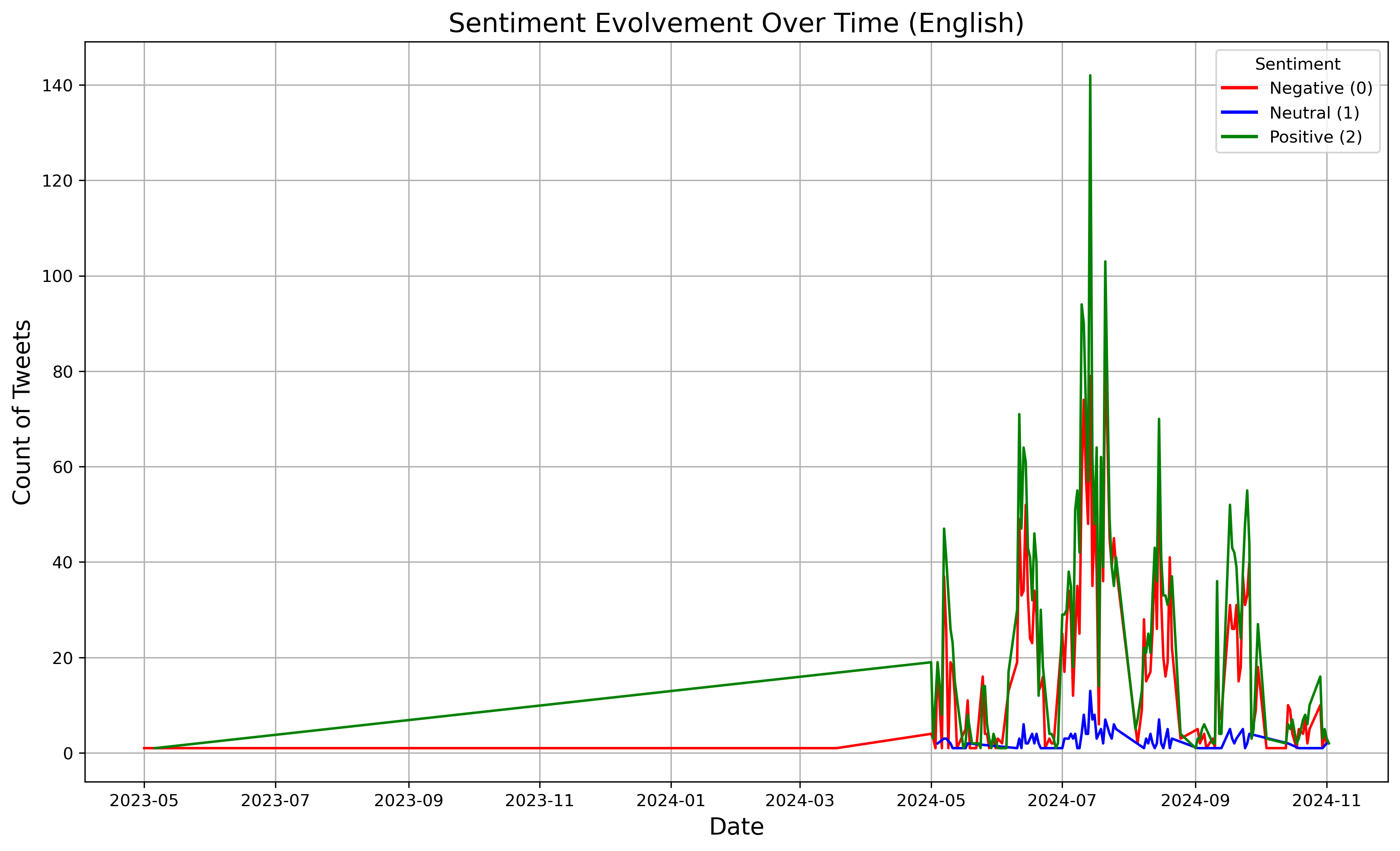}
    \caption{Sentiment trend for English IO drivers. Overall, positive and negative sentiments are close, except during August 2024 when positive sentiment reaches a pick much higher than negative sentiment. }
    \label{fig:en-sent-driver}
\end{figure}
\begin{figure}[H]
    \centering
    \includegraphics[width=\linewidth]{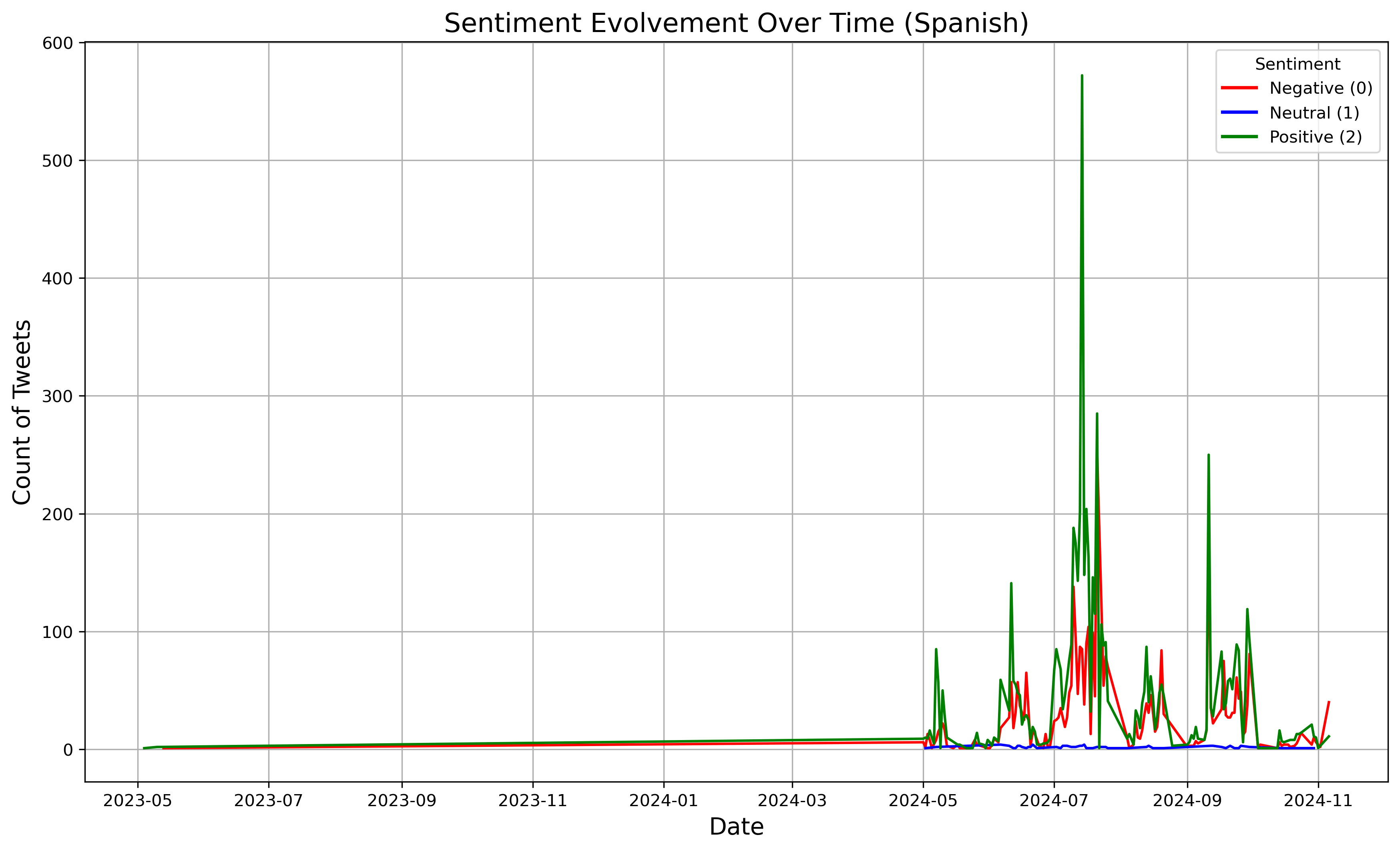}
    \caption{Sentiment trend for Spanish IO drivers. Overall, positive and negative sentiments are close, except during August 2024 when positive sentiment is clearly more frequent than negative sentiment.}
    \label{fig:es-sent-driver}
\end{figure}


\subsection*{Common Tags and Domains among Bilingual Posters}

\begin{table}[H]
\centering
\resizebox{0.6\columnwidth}{!}{%
\begin{tabular}{|c|c|}
\hline
\textbf{Top Domains} & \textbf{Frequency} \\ \hline
\multicolumn{2}{|c|}{\textbf{English Tweets}} \\ \hline
breitbart.com & 238 \\ \hline
foxnews.com & 232 \\ \hline
newsbreak.com & 119 \\ \hline
rawstory.com & 116 \\ \hline
rumble.com & 110 \\ \hline
\multicolumn{2}{|c|}{\textbf{Spanish Tweets}} \\ \hline
voz.us & 315 \\ \hline
clarin.com & 136 \\ \hline
cnn.com & 90 \\ \hline
inmigracionyvisas.com & 81 \\ \hline
rt.com & 66 \\ \hline
\end{tabular}%
}
\caption{Comparison of top 5 Domains in English and Spanish Tweets. English tweets prominently feature conservative media like \textit{breitbart.com} and \textit{foxnews.com}, while Spanish tweets focus on domains such as \textit{voz.us} and \textit{clarin.com}, reflecting differences in media consumption between languages.}
\label{tab:top_domains_bilingual}
\end{table}

\begin{table}[H]
\centering
\resizebox{0.5\columnwidth}{!}{%
\begin{tabular}{|c|c|}
\hline
\textbf{Top Tags} & \textbf{Frequency} \\ \hline
\multicolumn{2}{|c|}{\textbf{English Tweets}} \\ \hline
trump2024 & 1331 \\ \hline
maga & 1109 \\ \hline
trump & 345 \\ \hline
biden & 301 \\ \hline
bidenharris2024 & 116 \\ \hline
\multicolumn{2}{|c|}{\textbf{Spanish Tweets}} \\ \hline
trump2024 & 970 \\ \hline
maga & 540 \\ \hline
maripily & 382 \\ \hline
maripilyrivera & 361 \\ \hline
biden & 246 \\ \hline
\end{tabular}%
}
\caption{Comparison of top 5 Tags in English and Spanish Tweets. Both languages prominently feature \#trump2024 and \#maga, reflecting shared political themes, but Spanish tweets also include non-political tags like \#maripily and \#maripilyrivera, indicating a broader thematic focus.}
\label{tab:top_tags_bilingual}
\end{table}

\subsection*{Interpretation of High Eigenvector Centrality}
\label{subsec:high-centrality}
Upon analyzing nodes with the highest eigenvector centrality across multiple clusters and networks beyond Co-Hashtag Network, we found that most of these nodes corresponded to organic users with low Botometer scores \citep{10.1145/3340531.3412698}. This finding underscores the importance of caution when using centrality scores as a feature for detecting social bots. Although our study is based on a relatively small dataset and we did not conduct a large-scale Botometer analysis on all high-centrality nodes, our results suggest that solely relying on eigenvector centrality may lead to false positives in bot detection. Future research should adopt a more nuanced approach by combining centrality scores with additional features to improve detection accuracy.

\end{document}